\documentclass[nofootinbib,superscriptaddress,twocolumn,aps,prc,showpacs,10pt]{revtex4-1}

\usepackage{bm}
\usepackage{amsmath}
\usepackage{amssymb}
\usepackage{graphicx}
\usepackage{mathrsfs}
\usepackage{hyperref}
\usepackage{times}
\usepackage{color}

\begin{document}
\title{Clusterization and deformation of multi-$\Lambda$ hypernuclei within relativistic mean-field model}

\author{Yusuke Tanimura} \email{tanimura@nucl.phys.tohoku.ac.jp}
\affiliation{Department of Physics, Tohoku University, Sendai, 980-8578, Japan}
\affiliation{Graduate Program on Physics for the Universe, Tohoku University, Sendai, 980-8578, Japan}
\date{\today}

\begin{abstract}
Deformed multi-$\Lambda$ hypernuclei are studied within a relativistic mean-field model. 
In this paper, we take some $N=Z$ ``hyper isotope'' chains, {\it i.e.}, 
$^{8+n}_{\ \ n\Lambda}{\rm Be}$, $^{20+n}_{\ \ \ n\Lambda}{\rm Ne}$, 
and $^{28+n}_{\ \ \ n\Lambda}{\rm Si}$ systems 
where $n = 2$, $4$ for Be, and  $n = 2$, $8$ for Ne and Si. 
A sign of two-$^6_{2\Lambda}$He cluster structure is observed in the two-body correlation 
in $^{12}_{4\Lambda}$Be. 
In the Ne hyper isotopes, the deformation is slightly reduced by addition of $\Lambda$ hyperons 
whereas it is significantly reduced or even disappears in the Si hyper isotopes. 
\end{abstract}


\maketitle


\section{Introduction}

Hypernuclear physics studies have been made both theoretically and experimentally \cite{HT06,GHM16}. 
A primary motivation is the unified understanding of the baryon-baryon interaction based on the flavor SU(3) symmetry. 
The knowledge of general baryon-baryon interactions is essential for 
the property of neutron-star matter, in which the hyperons would emerge at high densities. 
Another importance of hypernuclear physics is that a hyperon could be a non-destructive probe of nuclear structure 
with the strong interaction since it is free from the Pauli principle of nucleons. 

In any physics case, precise information on hyperon-nucleon and hyperon-hyperon interactions is essential. 
However, it is difficult to perform nucleon-hyperon or hyperon-hyperon scattering experiments 
because of the short life times of hyperons. 
The main strategy is to produce hyperon(s) in an ordinary nucleus via 
$(K^-,\pi^-)$, $(\pi^+,K^+)$, $(e,e'K^+)$, or $(K^-,K^+)$ reactions 
to extract information on the interactions \cite{HT06,GHM16,Fr95,Naka10}. 
Most of the experimental data are limited to single-$\Lambda$ hypernuclei, and 
there are also a few data for double-$\Lambda$ and $\Xi$ nuclei 
\cite{Fr95, Naka10,Da63, Pr66,Nagara,Mikage, Mino, Yo17}.
Double-$\Lambda$ hypernuclei are of special importance since it provides information of hyperon-hyperon interaction. 

Besides the original motivations of hypernuclear physics, 
the property of nuclear system with one or more hyperons itself has been found to be interesting. 
In particular, so-called ``impurity effects'' of $\Lambda$ hyperon when a few of them are added to an 
ordinary nucleus have been of great interest. 
The impurity effects have been extensively studied with various theoretical models. 
The attractive $N$-$\Lambda$ and $\Lambda$-$\Lambda$ interactions 
and the absence of the Pauli principle 
lead to a variety of impurity effects in single- and double-$\Lambda$ nuclei, such as 
shrinkage of inter-cluster distance \cite{MBI83,Mo85,IBM85,HKMM99,HKYMR10,En17}, 
modification of deformation \cite{MH08,SMHS10,MHK11, Lu11, Is11,Is12, Is13,Cui17,Mei18}, 
extension of the drip line \cite{VPLR98}, 
emergence of new collective modes \cite{MH12}, 
change of fission barrier \cite{MCH09,MC11}, 
and existence of ``supersymmetric'' or ``genuine hypernuclear'' states \cite{DG76,MBI83}. 

The multi-$\Lambda$ system with more than two $\Lambda$ particles also is intriguing. 
We emphasize that hypernuclei with multiple strangeness are expected 
to be produced in future experiments \cite{Sa10} 
although it is not feasible at the moment to produce it experimentally. 
In such systems, the $\Lambda$ hyperon is no longer a small amount of ``impurity'', 
but it would be regarded as a third constituent of the nuclear many-body system, 
giving another dimension to the nuclear chart. 
The properties of the hyperon different from the nucleon would bring about nontrivial structures 
of nuclear system. 
As a matter of fact, 
multi-strangeness nuclei have been discussed in some theoretical calculations 
\cite{BIM, MIB83, IBM85, Rufa90, Sch92, MZ93, Sch94, LM02,FSV08,SS09,KMGR15,MKG17,GBKM18}. 
Those include the followings:  
$^6_{2\Lambda}$He clusters in hypernuclei \cite{BIM,MIB83,IBM85}, 
hyperon halo \cite{Rufa90, MZ93,Sch92,LM02}, 
competition between $\Lambda$ and $\Xi$ hyperons in finite nuclei \cite{MKG17,Sch94}, 
shell structure \cite{Sch94}, 
dense hypernuclei \cite{Sch94}, 
binding energy systematics \cite{KMGR15}, 
$\Lambda$-$\Lambda$ pairing correlation \cite{GBKM18}, 
and stability of light multi-strange nuclei \cite{FSV08,SS09}. 
So far, to the best of our knowledge, the calculations have been limited to spherical systems except for 
the ones in Refs. \cite{BIM}, \cite{MIB83}, and \cite{SS09}, 
where a microscopic cluster model, a molecular orbital model, and a cluster model were applied 
to the Be hyper isotopes, respectively. 
In Ref. \cite{MIB83}, it was demonstrated that the $\Lambda$ hyperons in $^{12}_{4\Lambda}$Be 
favor a localization around the two $\alpha$ clusters. 

In this paper, we focus on the clusterization and deformation properties of multi-$\Lambda$ systems. 
We employ a relativistic mean field (RMF) model, 
without any assumption on spatial symmetry and cluster structure. 
The RMF is one of the widely used models for hypernuclei (see Ref. \cite{HY16} for a review). 
We take the following $N=Z$ hyper isotope chains: 
$^{8+n}_{\ \ n\Lambda}{\rm Be}$, $^{20+n}_{\ \ \ n\Lambda}{\rm Ne}$, 
and $^{28+n}_{\ \ \ n\Lambda}{\rm Si}$ systems 
where $n = 2$, $4$ for Be, and  $n = 2$, $8$ for Ne and Si. 
In particular, we make an analysis on the cluster structure using the two-body 
correlation as well as the one-body density distribution. 

The paper is organized as follows. In Sec. \ref{sec:model}, we introduce the RMF model employed 
in this work. 
In Sec. \ref{sec:results1}, we present the results of the Be hyper isotopes and discuss 
their structures with a special attention to the cluster structure. 
In Sec. \ref{sec:results2}, we discuss the deformations of the Ne and Si hyper isotopes. 
In Sec. \ref{sec:summary}, we give the summary and perspectives.

\section{Model and numerical details}\label{sec:model}

To describe the multi-$\Lambda$ nucleus, we use a meson-exchange model 
with non-linear couplings for RMF theory. 
The Lagrangian density is given by 
\begin{eqnarray}
{\cal L} &=& 
\bar\psi_N(i\partial\!\!\!/-m_N)\psi_N
+\bar\psi_\Lambda(i\partial\!\!\!/-m_\Lambda)\psi_\Lambda
\nonumber \\
&&
+\frac{1}{2}(\partial_\mu\sigma)(\partial^\mu\sigma) - \frac{1}{2}m_\sigma^2\sigma^2
-\frac{c_3}{3}\sigma^3-\frac{c_4}{4}\sigma^4
\nonumber \\
&&
-\frac{1}{4}G^{\mu\nu}G_{\mu\nu} + \frac{1}{2}m_\omega^2\omega^\mu\omega_\mu
+\frac{d_4}{4}(\omega^\mu\omega_\mu)^2
\nonumber \\
&&
-\frac{1}{4}\vec R^{\mu\nu}\cdot\vec R_{\mu\nu}
+ \frac{1}{2}m_\rho^2\vec\rho^\mu\cdot\vec\rho_\mu
\nonumber \\
&&
 -\frac{1}{4}F^{\mu\nu}F_{\mu\nu}
\nonumber \\
&&
-\bar\psi_N\left(g_{\sigma N}\sigma+ g_{\omega N}\omega\!\!\!/ 
+g_{\rho N}\vec\rho\!\!\!/\cdot\vec\tau 
+eA\!\!\!/\frac{1-\tau_3}{2}
\right)\psi_N
\nonumber \\
&&
-\bar\psi_\Lambda\left(g_{\sigma\Lambda}\sigma
+ g_{\omega\Lambda}\omega\!\!\!/ 
+\frac{f_{\omega\Lambda}}{4m_\Lambda}G_{\mu\nu}\sigma^{\mu\nu}
\right)\psi_\Lambda, 
\label{eq:L}
\end{eqnarray}
where $\psi_N$ and $\psi_\Lambda$ are nucleon and $\Lambda$ hyperon fields, and 
$G^{\mu\nu}=\partial^\mu \omega^\nu- \partial^\nu  \omega^\mu$, 
$\vec R^{\mu\nu}=\partial^\mu \vec\rho^\nu- \partial^\nu\vec\rho^\mu$, and
$F^{\mu\nu}=\partial^\mu A^\nu- \partial^\nu A^\mu$ 
are the field tensors of the vector mesons $\omega$ and $\rho$, and the photon, respectively. 
$\vec\tau$ is the Pauli matrix in the isospin space. 
We take PK1 parameter set \cite{pk1} for the nucleon-meson couplings 
and a parameter set in Ref. \cite{pk1y} for $\Lambda$-meson couplings. 
In Ref. \cite{pk1y}, $g_{\sigma\Lambda}$ is fitted to the binding energy of $^{40}_\Lambda$Ca ($g_{\sigma\Lambda}=0.618g_{\sigma N}$). 
The naive quark model value $g_{\omega\Lambda}=(2/3)g_{\omega N}$ is taken for 
the Yukawa coupling between $\Lambda$ and $\omega$. 
As for the tensor $\omega-\Lambda$ coupling, $f_{\omega\Lambda}=-g_{\omega\Lambda}$ 
from a quark model \cite{CW91} is used as in Ref. \cite{MJ94}. 
The tensor coupling modifies the spin-orbit potential and 
reduces the spin-orbit splittings of the single-particle energies of $\Lambda$ hyperon \cite{MJ94,pk1y,TH12}. 
The $\Lambda$ hyperon mass $m_\Lambda$ is taken to be $1115.6$ MeV. 
This model with the parameters thus determined reproduces the observed binding  
energies of light to heavy single-$\Lambda$ hypernuclei reasonably well \cite{MJ94,pk1y}. 
In the present study, we do not take into account the pairing correlations among nucleons and $\Lambda$ hyperons. 

\begin{widetext}

Here we make a remark on the $\Lambda$-$\Lambda$ interaction. 
There have been a discussion that the $\sigma$-$\omega$ RMF model underestimates the observed 
double-$\Lambda$ binding energy $B_{\Lambda\Lambda}$ and 
$\Lambda$-$\Lambda$ interaction energy $\Delta B_{\Lambda\Lambda}$ \cite{Sch94,MLM98}. 
$B_{\Lambda\Lambda}$ and $\Delta B_{\Lambda\Lambda}$ are 
defined as follows, 
\begin{eqnarray}
B_{\Lambda\Lambda}(^A_{2\Lambda}Z) &=& B(^{A-2}Z)-B(^A_{2\Lambda}Z), 
\label{eq:BLL}
\\
\Delta B_{\Lambda\Lambda}(^A_{2\Lambda}Z) &=& B_{\Lambda\Lambda}(^A_{2\Lambda}Z)
-2B_{\Lambda}(^{A-1}_{\ \ \ \ \Lambda}Z), 
\label{eq:dBLL}
\end{eqnarray}
where $B(^AZ)$ is the binding energy of a nucleus $^AZ$, 
and $B_{\Lambda}(^{A-1}_{\ \ \ \ \Lambda}Z) = B(^{A-1}_{\ \ \ \ \Lambda}Z)-B(^{A-2}Z)$. 
In Refs. \cite{Sch94} and \cite{MLM98}, the discrepancies were attributed to contributions from strange 
$\sigma^*$ and $\phi$ mesons which are coupled only to $\Lambda$ hyperon. 
However, the discussion was based on the old data \cite{Pr66,Da63,Fr95} 
for $B_{\Lambda\Lambda}$ and $\Delta B_{\Lambda\Lambda}$, 
whereas the $B_{\Lambda\Lambda}$ and $\Delta B_{\Lambda\Lambda}$ values in newer data are considerably smaller \cite{Naka10}.
In Table \ref{tb:BLL} are listed the values of $B_{\Lambda\Lambda}$ and 
$\Delta B_{\Lambda\Lambda}$ obtained with 
the present RMF model in comparison with the experimental values (see also Table I in Ref. \cite{MLM98}). 
We see that the RMF results agree better with the newer data. 
Nevertheless, it has to be noted that the effect of beyond-mean-field correlation that could be 
significant in the light systems is missing in RMF 
although there is a partial cancellation of the correlation energies upon the subtraction in Eq. (\ref{eq:BLL}). 
Moreover, the experimental $B_{\Lambda\Lambda}$ and $\Delta B_{\Lambda\Lambda}$ 
values were deduced under some assumption 
on the formation and decay processes of the double-$\Lambda$ nuclei \cite{Naka10}. 
Therefore, we leave discussions on extra $\Lambda$-$\Lambda$ interaction and correlation for future works 
and adopt the phenomenological model in Eq. (\ref{eq:L}) fitted only to a single-$\Lambda$ nucleus. 
More of new data from the ongoing analyses of J-PARC E-07 experiment are awaited \cite{Mino,Yo17}. 

\begin{table}
\caption{The double-$\Lambda$ binding energies $B_{\Lambda\Lambda}$ and 
$\Lambda$-$\Lambda$ interaction energies $\Delta B_{\Lambda\Lambda}$
obtained with the present RMF model. 
For comparison, the experimental data in Refs. \cite{Naka10,Fr95,Mino} are also listed. 
Note that the values of $B_{\Lambda\Lambda}$ and $\Delta B_{\Lambda\Lambda}$ 
for $^6_{2\Lambda}$He from Ref. \cite{Naka10} are given by the weighted average of 
NAGARA event ($B_{\Lambda\Lambda}=6.91\pm0.16$ MeV and 
$\Delta B_{\Lambda\Lambda}=0.67\pm0.17$ MeV) \cite{Nagara}
and MIKAGE event ($B_{\Lambda\Lambda}=10.06\pm1.72$ MeV and 
$\Delta B_{\Lambda\Lambda}=3.82\pm1.72$ MeV) \cite{Mikage}. 
The former have been uniquely identified as $^6_{2\Lambda}$He \cite{Naka10,Nagara}
whereas the latter was only found most probable to be $^6_{2\Lambda}$He \cite{Naka10,Mikage}. 
The data of Ref. \cite{Mino} are the very recent results of MINO event in J-PARC E-07 experiment, 
the candidates of which are $^{10}_{2\Lambda}$Be, $^{11}_{2\Lambda}$Be (most likely), and 
$^{12}_{2\Lambda}$Be. }
\begin{tabular}{ccccccccc}
\hline\hline
                                      & \multicolumn{4}{c}{ $B_{\Lambda\Lambda}$ (MeV)}
                                      & \multicolumn{4}{c}{ $\Delta B_{\Lambda\Lambda}$ (MeV)}\\
                                      & RMF   & exp. \cite{Mino} & exp. \cite{Naka10} & exp. \cite{Fr95}
                                      & RMF   & exp. \cite{Mino} & exp. \cite{Naka10} & exp. \cite{Fr95}\\
\hline
$^6_{2\Lambda}$He      & $4.24$   & $-$ & $6.93 \pm 0.16$    & $10.9\pm 0.6$ 
                                         & $1.08$   & $-$ &  $0.70 \pm 0.17$    & $4.7\pm 0.6$ \\ 
$^{10}_{2\Lambda}$Be & $13.23$ & $(15.05\pm 0.11)$ &  $11.90 \pm 0.13$   & $17.7\pm 0.4$
                                         & $0.59$   & $(1.63\pm 0.14)$ &  $-1.52 \pm 0.15$    & $4.3\pm 0.4$ \\ 
$^{13}_{2\Lambda}$B  & $-$         & $-$ &  $23.3 \pm 0.7$   & $27.5\pm 0.7$ 
                                         & $-$         & $-$ &  $0.6 \pm 0.8$    & $4.8\pm 0.7$ \\ 
\hline\hline
\end{tabular}
\label{tb:BLL}
\end{table}
\end{widetext}

The model Lagrangian given above is solved within the mean-field and the no-sea approximations. 
We assume time-reversal invariance and charge conservation of the mean-field state, 
{\it i.e.}, the time-odd or charged vector fields vanish.  
The only non-zero components are their time-like and neutral components, 
$\omega^0$, $\rho^0_3$, and $A^0$, 
where the subscript $3$ on the $\rho$ meson field means the third component in the isospace. 

The scalar-isoscalar density $\rho_S$, vector-isoscalar and -isovector densities $j^0$ and $j_3^0$ 
of nucleon are defined in terms of the single-particle wave functions of nucleon $\psi_k^{(N)}$ as 
\begin{eqnarray}
\rho_S(\bm r) &=& \sum_{k\in{\rm occ}}\bar\psi_k^{(N)}(\bm r)\psi_k^{(N)}(\bm r),
\label{eq:rhos}
\\
j^0(\bm r) &=& \sum_{k\in{\rm occ}}\bar\psi_k^{(N)}(\bm r)\gamma^0\psi_k^{(N)}(\bm r), 
\\
j^0_3(\bm r) &=& \sum_{k\in{\rm occ}}\bar\psi_k^{(N)}(\bm r)\gamma^0\tau_3\psi_k^{(N)}(\bm r), 
\end{eqnarray}
where $k$ runs over the occupied nucleon states. 
The scaler, vector, and tensor densities of $\Lambda$ hyperon are defined in terms of the single-particle 
wave function of $\Lambda$ hyperon $\psi_k^{(\Lambda)}$ as
\begin{eqnarray}
\rho_{S\Lambda}(\bm r) &=& \sum_{k\in{\rm occ}}\bar\psi_k^{(\Lambda)}(\bm r)\psi_k^{(\Lambda)}(\bm r),
\\
j^0_\Lambda(\bm r) &=& \sum_{k\in{\rm occ}}\bar\psi_k^{(\Lambda)}(\bm r)\gamma^0\psi_k^{(\Lambda)}(\bm r), 
\\
\bm V_{T\Lambda}(\bm r) &=&
\sum_{k\in{\rm occ}}\bar\psi_k^{(\Lambda)}(\bm r)i\bm\alpha\psi_k^{(\Lambda)}(\bm r), 
\end{eqnarray}
where $k$ here runs over the occupied $\Lambda$ hyperon states. 

The equations of motion for the meson and electromagnetic fields read 
\begin{eqnarray}
(-\bm\nabla^2+m_\sigma^2)\sigma &=& -g_{\sigma N}\rho_S-c_3\sigma^2 - c_4\sigma^3
\nonumber \\
&&
-g_{\sigma\Lambda}\rho_{S\Lambda}, 
\\
(-\bm\nabla^2+m_\omega^2)\omega^0 &=&
 g_{\omega N} j^0 - d_4\left(\omega^0\right)^3
\nonumber \\
&&
+g_{\omega\Lambda}j^0_\Lambda
+\frac{f_{\omega\Lambda}}{2m_\Lambda}\bm\nabla\cdot\bm V_{T\Lambda}, 
\\
(-\bm\nabla^2+m_\rho^2)\rho^0_3  &=& g_{\rho N} j^0_3,
\\
-\bm\nabla^2A^0 &=& \frac{e}{2}(j^0-j^0_3). 
\end{eqnarray}
Notice that there are the contributions from the $\Lambda$ hyperon to the 
sources of the $\sigma$ and $\omega$ fields.

The Dirac equation for the nucleon single-particle wave function is given by 
\begin{equation}
\left[-i\bm\alpha\cdot\bm\nabla + V_N + \beta(m_N+S_N)\right]\psi_k^{(N)}
=\epsilon_k\psi_k^{(N)}, 
\end{equation}
where 
\begin{eqnarray}
S_N &=& g_{\sigma N}\sigma, \\
V_N &=& g_{\omega N}\omega^0 + g_{\rho N}\rho^0_3\tau_3 + eA^0\frac{1-\tau_3}{2}. 
\end{eqnarray}
Note that $\tau_3=+1$ for neutron and $\tau_3=-1$ for proton. 
The Dirac equation for the single-particle wave function of $\Lambda$ hyperon is given by 
\begin{equation}
\left[-i\bm\alpha\cdot\bm\nabla + V_\Lambda + \beta(m_\Lambda+S_\Lambda)
+T_\Lambda \right]\psi_k^{(\Lambda)}
=\epsilon_k\psi_k^{(\Lambda)}, 
\end{equation}
where 
\begin{eqnarray}
S_\Lambda &=& g_{\sigma \Lambda}\sigma, \\
V_\Lambda &=& g_{\omega \Lambda}\omega^0, \\ 
T_\Lambda &=& \frac{f_{\omega\Lambda}}{2m_\Lambda}
i\beta\bm\alpha\cdot\left(-\bm\nabla\omega^0\right). 
\label{eq:potl}
\end{eqnarray}

The set of non-linear equations, Eqs. (\ref{eq:rhos}-\ref{eq:potl}), are solved self-consistently. 
In our numerical calculations, 
the single-particle wave functions are represented on 3-dimensional lattice in the real space \cite{THL15}. 
To avoid the fermion doubling, the derivative in the Dirac equation is computed 
in the momentum space with the fast Fourier transform \cite{Ren17}, for which we use FFTW library \cite{fftw}. 
The damped gradient iteration technique \cite{RC82,BLMR92} is used to solve the self-consistent 
mean-field equations (see also Ref. \cite{BSUR89}). 
The Klein-Gordon equations for the mesons and the Poisson equation for the Coulomb field are 
solved in the momentum space with the fast Fourier transform. 
For solving the Poisson equation in the momentum space, we employ the same method as in 
Refs. \cite{Sky3D,Poisson}. 
Calculations are performed with $24^3$ mesh points and lattice spacing of $0.8$ fm.

\section{Results and discussion: BE hyper isotopes}\label{sec:results1}

First, we show the results of 
$^{8}$Be, $^{10}_{2\Lambda}$Be, $^{12}_{4\Lambda}$Be.

In the density plots in this paper, we will only show the neutron and $\Lambda$ densities 
because the neutron and proton densities are almost the same in the $N=Z$ hyper isotopes.  
From here on, the neutron and $\Lambda$ densities will be referred to as 
$\rho_n$ and $\rho_\Lambda$, respectively,  
\begin{eqnarray}
\rho_n &\equiv& \frac{1}{2}(j^0+j^0_3), \\
\rho_\Lambda &\equiv& j^0_\Lambda. 
\end{eqnarray}

\begin{figure}
\begin{center}
\includegraphics[width=8.5cm]{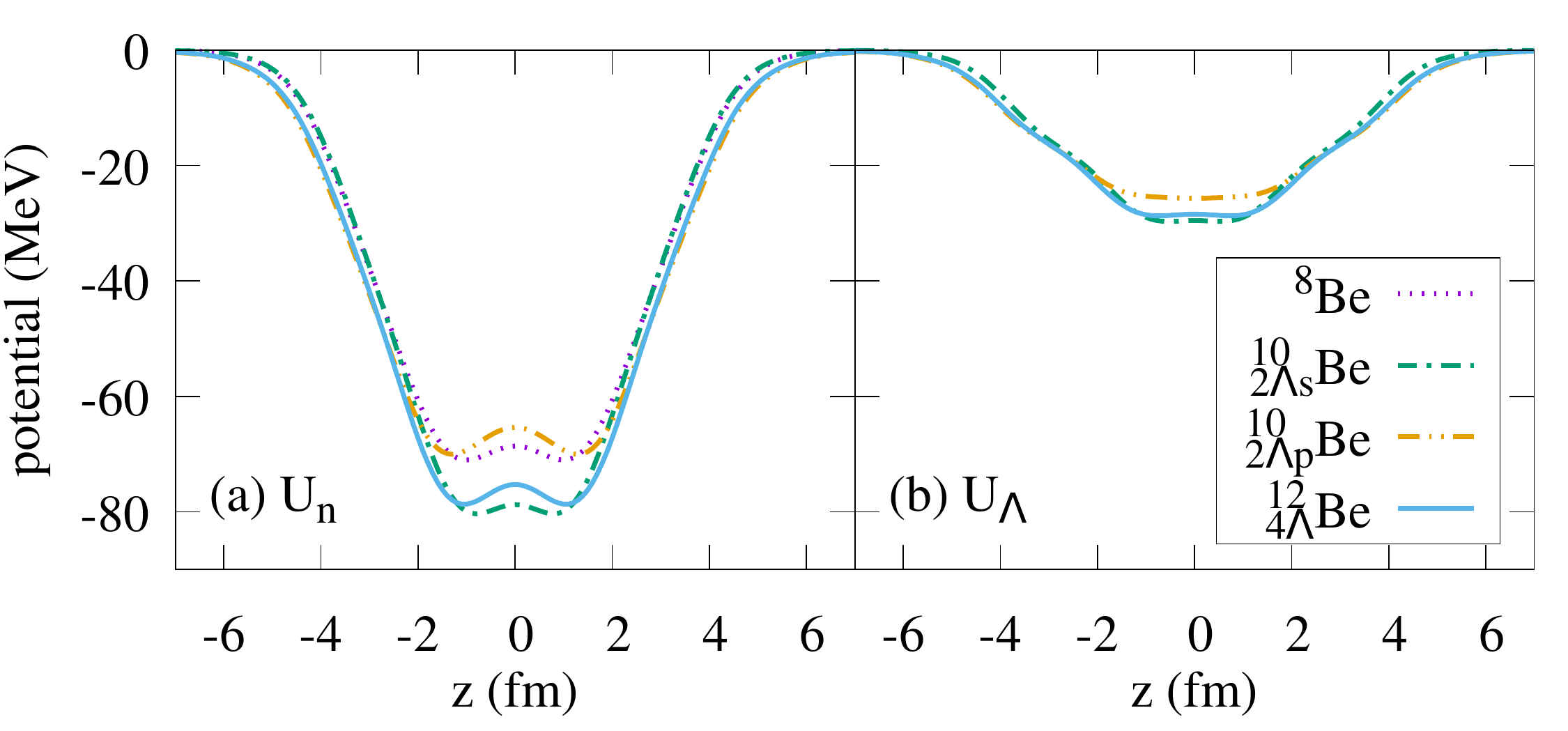}
\end{center}
\caption{Central potentials of (a) neutron and (b) $\Lambda$ hyperon of Be hyper isotopes, 
$^{8}$Be,  $^{10}_{2\Lambda s}$Be,  $^{10}_{2\Lambda p}$Be, and $^{12}_{4\Lambda}$Be 
along the symmetry axis. 
In $^{10}_{2\Lambda s}$Be and $^{10}_{2\Lambda p}$Be, the two $\Lambda$ hyperons occupy 
the $s$ and the $p$ orbital, respectively. 
}
\label{fig:pot_be}
\end{figure}

\begin{widetext}

In Fig. \ref{fig:den2d_be}, the density distributions of neutron and $\Lambda$ hyperon are presented. 
The neutron densities $\rho_n$ are drawn by contour map, while the $\Lambda$ densities $\rho_\Lambda$ 
are drawn by color map. 
Fig. \ref{fig:den2d_be} (a) shows the ground state of $^8$Be, 
and Fig. \ref{fig:den2d_be} (b) and (c) show 
the ground state of $^{10}_{2\Lambda}$Be with the two $\Lambda$ hyperons in the $s$ orbital 
(denoted as $^{10}_{2\Lambda s}$Be) and 
$^{10}_{2\Lambda}$Be with the two $\Lambda$ hyperons in the $p$ orbital (denoted as $^{10}_{2\Lambda p}$Be), respectively 
, and 
Fig. \ref{fig:den2d_be} (d) shows the ground state of $^{12}_{4\Lambda}$Be. 
$^{10}_{2\Lambda p}$Be is obtained by having the $\Lambda$'s occupy the second lowest 
single-particle level during the self-consistent iterations. 
In Fig. \ref{fig:pot_be} we show the central mean-field potentials of neutron and $\Lambda$, 
\begin{eqnarray}
U_n &\equiv& V_n + S_n, \\
U_\Lambda &\equiv& V_\Lambda + S_\Lambda. 
\end{eqnarray}
In Fig. \ref{fig:pot_be}, The dotted line represents $^8$Be, 
dot-dashed and double-dot-dashed lines represent $^{10}_{2\Lambda s}$Be 
and $^{10}_{2\Lambda p}$Be, respectively, and the solid line represents $^{12}_{4\Lambda}$Be.

\subsection{Structures of $^{10}_{2\Lambda}$Be with $\Lambda$'s in $s$ or $p$ orbitals}

Here we compare the structures of $^8$Be, $^{10}_{2\Lambda s}$Be, and $^{10}_{2\Lambda p}$Be. 
Note that the two $\Lambda$ hyperons in $^{10}_{2\Lambda p}$Be occupy the $p$ state along the 
symmetry axis of the system [Fig. \ref{fig:den2d_be} (c)]. 
Such a configuration
does not have an analog state in the non-strange isobars, {\it e.g.}, 
$^{10}$Be where two neutrons are added to $^8$Be instead of $\Lambda$ hyperons \cite{HT06,MBI83,Is11,Mei14,DG76}. 
Thus this state is unique in the hypernucleus and 
is called ``supersymmetric'' or ``genuine hypernuclear'' state \cite{MBI83,DG76}. 

We see in Fig. \ref{fig:den2d_be} (a)-(d) that the nucleon distributions 
in the Be hyper isotopes exhibit a well-developed $\alpha$ cluster structure. 
In $^{10}_{2\Lambda s}$Be, the $\Lambda$ particles stay around the middle between the two $\alpha$'s 
[Fig. \ref{fig:den2d_be} (b)]. Accordingly, the central potential for neutron, 
as seen from the dot-dashed curve in Fig. \ref{fig:pot_be} (a), becomes deeper at the middle due to the additional attraction 
by the $\Lambda$ hyperons. The nucleons are attracted toward the middle of the system, 
which leads to a slight reduction of the distance between the two $\alpha$ clusters. 
On the other hand, in $^{10}_{2\Lambda p}$Be in which the two $\Lambda$ hyperons are 
in the $p$ orbital, the potential becomes shallower and wider [double-dot-dashed curve Fig. \ref{fig:pot_be} (a)], 
and the nucleon distribution is stretched along the $z$ axis, which is the axis of symmetry. 
These effects are seen in the quadrupole deformation parameter $\beta_2$, 
the root-mean-squared (rms) radius $R$ of the system, and the $\alpha$-$\alpha$ distance 
$D_{\alpha\alpha}$, 
which are summarized in Table \ref{tb:q2r_be}. 
The quadrupole deformation parameter and the rms radius are defined as 
\begin{equation}
\beta_2 = \frac{4\pi}{5}\frac{\int d^3r\ r^2Y_{20}\rho}{\int d^3r\ r^2\rho}, 
\label{eq:b2}
\end{equation}
and
\begin{equation}
R = \sqrt{\frac{\int d^3r\ r^2\rho}{\int d^3r\ \rho}}, 
\label{eq:rmsr}
\end{equation}
respectively, where $\rho$ is the density of the particle that is considered. 
The $\alpha$-$\alpha$ distance $D_{\alpha\alpha}$ is defined as the distance between the two maxima 
of the nucleon density distribution along $z$ axis. 
One observes that 
$\beta_{2N}(^{10}_{\Lambda s}{\rm Be})<\beta_{2N}(^8{\rm Be})<\beta_{2N}(^{10}_{\Lambda p}{\rm Be})$, 
and the same relations for $R_{N}$ and $D_{\alpha\alpha}$ as well. 
We note that a similar result was obtained also in Ref. \cite{Is11} with 
an antisymmetrized molecular dynamics model for $^{9}_\Lambda$Be when a $\Lambda$ particle 
occupies either $s$ or $p$ orbital.

\begin{figure}
\begin{center}
\includegraphics[width=18.cm]{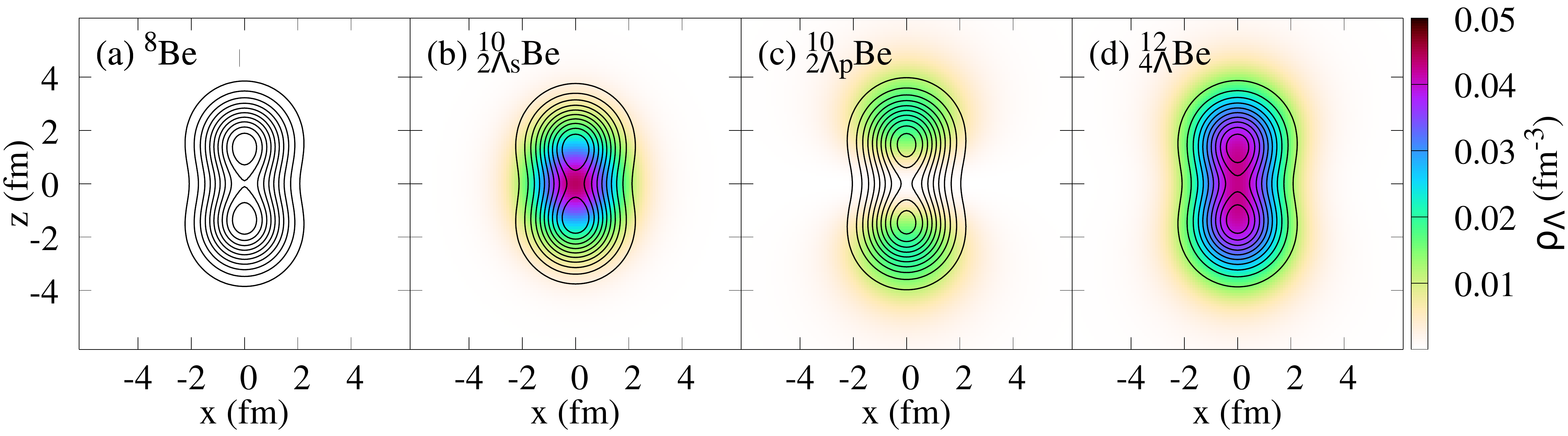}
\end{center}
\caption{Neutron and $\Lambda$ hyperon density distributions of 
(a) $^{8}$Be, (b) $^{10}_{2\Lambda}$Be with $\Lambda$'s in $s$ orbital ($^{10}_{2\Lambda s}$Be), 
(c) $^{10}_{2\Lambda}$Be with $\Lambda$'s in $p$ orbital ( $^{10}_{2\Lambda p}$Be), and 
(d) $^{12}_{4\Lambda}$Be. 
The neutron densities $\rho_n$ are shown by contour map, which starts from 0.1 fm$^{-3}$ 
with the increments by 0.1 fm$^{-3}$. 
The $\Lambda$ hyperon densities  $\rho_\Lambda$ are shown by color map. 
The densities at $y = 0$ as a function of $z$ and $x$ are plotted. 
$z$ (horizontal) is the symmetry axis. }
\label{fig:den2d_be}
\end{figure}

\end{widetext}

\begin{table}
\caption{Quadrupole deformation parameters $\beta_{2}$, root-mean-squared radii $R$, 
and the distance between the two $\alpha$'s $D_{\alpha\alpha}$ of 
the Be hyper isotope chain. $\beta_{2N}$ and $R_N$ are deformation parameter and radius 
calculated from the nucleon density, 
and $\beta_{2\Lambda}$ and $R_\Lambda$ are calculated from the $\Lambda$ hyperon density. 
$N_\Lambda$ is the number of $\Lambda$ hyperons. 
$\Lambda_s$ and $\Lambda_p$ for $N_\Lambda = 2$ mean that the $\Lambda$ particles are 
in $s$ and $p$ orbitals, respectively. }
\begin{tabular}{lccccc}
\hline\hline
$N_\Lambda$ & $\beta_{2N}$ & $\beta_{2\Lambda}$ & $R_N$ (fm) & $R_\Lambda$ (fm) & 
$D_{\alpha\alpha}$ (fm)\\
\hline
\multicolumn{6}{c}{Be hyper isotopes} \\
$0$ &                    $0.67$ & $-$       & $2.43$ & $-$ & $2.8$\\
$2\Lambda_s$ &  $0.64$ & $0.18$ & $2.36$ & $2.59$  & $2.6$\\
$2\Lambda_p$ &  $0.72$ & $0.71$ & $2.49$ & $4.27$  & $3.0$\\
$4$ &                     $0.68$ & $0.57$ & $2.42$ & $3.49$  & $2.8$\\
\hline\hline
\end{tabular}
\label{tb:q2r_be}
\end{table}

\subsection{Clusterization in $^{12}_{4\Lambda}$Be}

\begin{figure}
\begin{center}
\includegraphics[width=8.5cm]{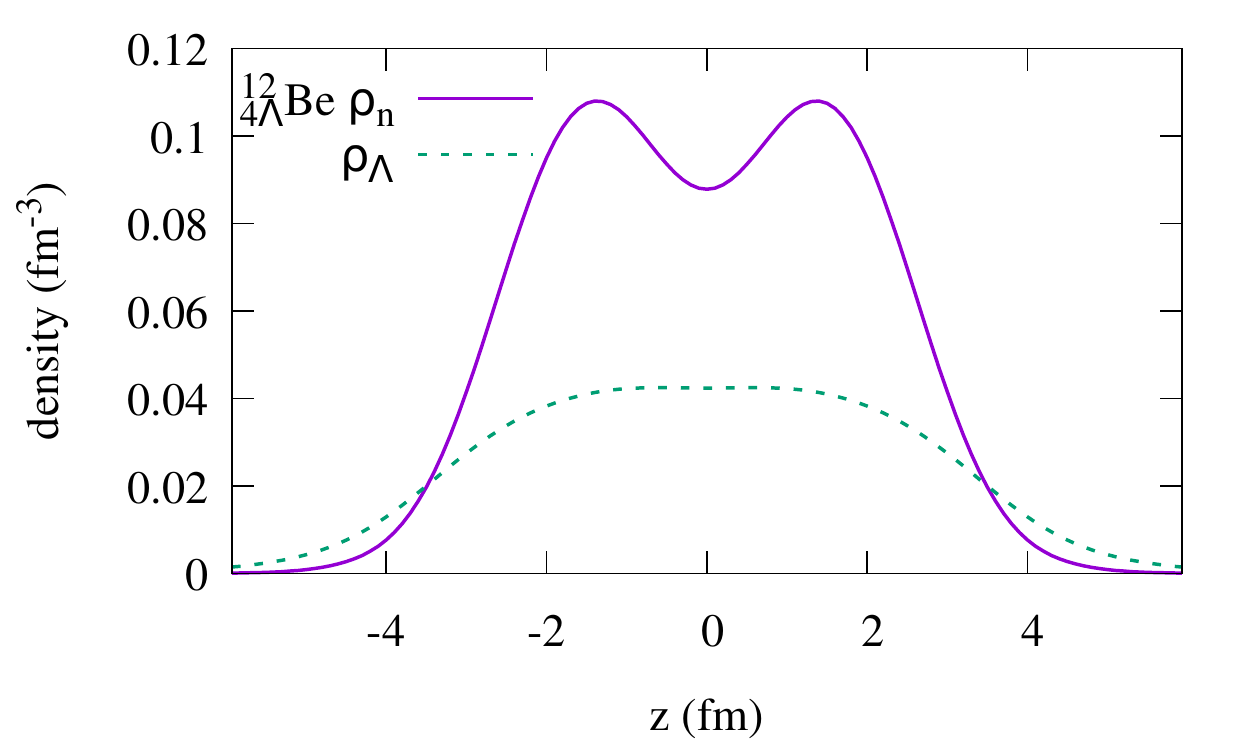}
\end{center}
\caption{Comparison between the neutron (solid curve) and $\Lambda$ hyperon (dashed curve) 
densities in $^{12}_{4\Lambda}$Be along the symmetry axis. }
\label{fig:den1d_be}
\end{figure}

In $^{12}_{4\Lambda}$Be with two more $\Lambda$'s, the density of $\Lambda$ hyperon 
becomes strongly deformed [Fig. \ref{fig:den2d_be} (d)], 
and the nucleon recover nearly the same deformation, radius, and $\alpha$-$\alpha$ distance as in the $^8$Be normal isotope 
(see Table \ref{tb:q2r_be}). 
However, as we see in Fig. \ref{fig:den1d_be}, the $\Lambda$ hyperon density distribution 
does not have a neck at the middle as the nucleon density distribution does. 
An interpretation to this behavior of the $\Lambda$ hyperon density is 
that the $\Lambda$'s are not as tightly bound to $\alpha$ 
as the nucleons are, so they can move more freely between the two clusters. 
As a consequence, the density distribution at the mean-field level becomes flat at the middle. 

We make a further analysis on the cluster structure in $^{12}_{4\Lambda}$Be, 
using the fermion localization function \cite{BE90,RMUO11,ZSN16,SN17,EKN17,JSSN18}. 
The localization function is a measure of localization, 
which is related to the spatial two-body correlation between two like-spin fermions of the same kind. 
The localization function for a particle of kind $q=n$, $p$, or $\Lambda$ with 
spin $\sigma=+1/2\ (\uparrow)$ or $-1/2\ (\downarrow)$ is defined as 
\begin{equation}
{\cal C}_{q\sigma}(\bm r) = 
\left[1+\left(\frac{\rho_{q\sigma}\tau_{q\sigma}-\bm j_{q\sigma}^2-\frac{1}{4}(\bm\nabla\rho_{q\sigma})^2}
{\rho_{q\sigma}\tau_{q\sigma}^{\rm TF}}\right)^2\right]^{-1}. 
\label{eq:C}
\end{equation}
Here, $\tau^{\rm TF}_{q\sigma} = \frac{3}{5}(6\pi^2)^{2/3}\rho_{q\sigma}^{5/3}$ is the 
Thomas-Fermi kinetic energy density, and 
\begin{eqnarray}
\tau_{q\sigma}(\bm r) &=& \frac{1}{2}\sum_{k\in q,{\rm occ}} [\bm\nabla\psi_k^\dagger(\bm r)]\cdot
(1+2\sigma\Sigma^3)[\bm\nabla\psi_k(\bm r)], 
\end{eqnarray}
and 
\begin{eqnarray}
\bm j_{q\sigma}(\bm r) &=& 
{\rm Im}\left[\sum_{k\in q,{\rm occ}}
\psi_k^\dagger(\bm r)(1+2\sigma\Sigma^3)\bm\nabla\psi_k(\bm r)\right], 
\end{eqnarray}
where $\Sigma^3=\left(\begin{array}{cc}
\sigma_z & 0\\
0 & \sigma_z 
\end{array}\right)$ is the spin operator, 
and $k$ runs over the occupied states of the fermion kind $q$.
Note that $\bm j_{q\sigma} = \bm 0$, $\rho_{q\uparrow} = \rho_{q\downarrow}$, 
and $\tau_{q\uparrow} = \tau_{q\downarrow}$
in the present case with the time-reversal symmetry. 

A value of ${\cal C}_{q\sigma}$ close to one is the sign of localization \cite{RMUO11}, 
which means that the probability of finding two particles with the same spin close to each other is very low. 
${\cal C}_{q\sigma}\approx 1$ simultaneously for all the spin-isospin combinations is 
a minimal necessary condition of $\alpha$ clusterization \cite{RMUO11}. 
The $\alpha$ cluster correlation in terms of the localization function can be naturally extended to 
$^6_{2\Lambda}$He:  
${\cal C}_{q\sigma}\approx 1$ at the same spacial region for 
$n\uparrow$, $n\downarrow$, $p\uparrow$, $p\downarrow$, 
$\Lambda\uparrow$, and $\Lambda\downarrow$ implies $^6_{2\Lambda}$He clusterization. 
In the present case with $N=Z$ and time-reversal symmetry, the wave functions of 
neutron and proton are approximately the same, and the spin-up and -down components are exactly the same, 
so it suffices to consider only the neutron spin-up and $\Lambda$ spin-up components. 

Since the localization is not a meaningful quantity in the regions where the one-body density is close to zero, 
we look at the localization function multiplied by the normalized one-body density
\begin{equation}
\overline{\cal C}_{q\sigma}(\bm r) = {\cal C}_{q\sigma}(\bm r)
\frac{\rho_{q\sigma}(\bm r)}{{\rm max}\rho_{q\sigma}(\bm r)}, 
\label{eq:Cbar}
\end{equation}
as was done in Ref. \cite{ZSN16}.

\begin{figure}
\begin{center}
\includegraphics[width=8.5cm]{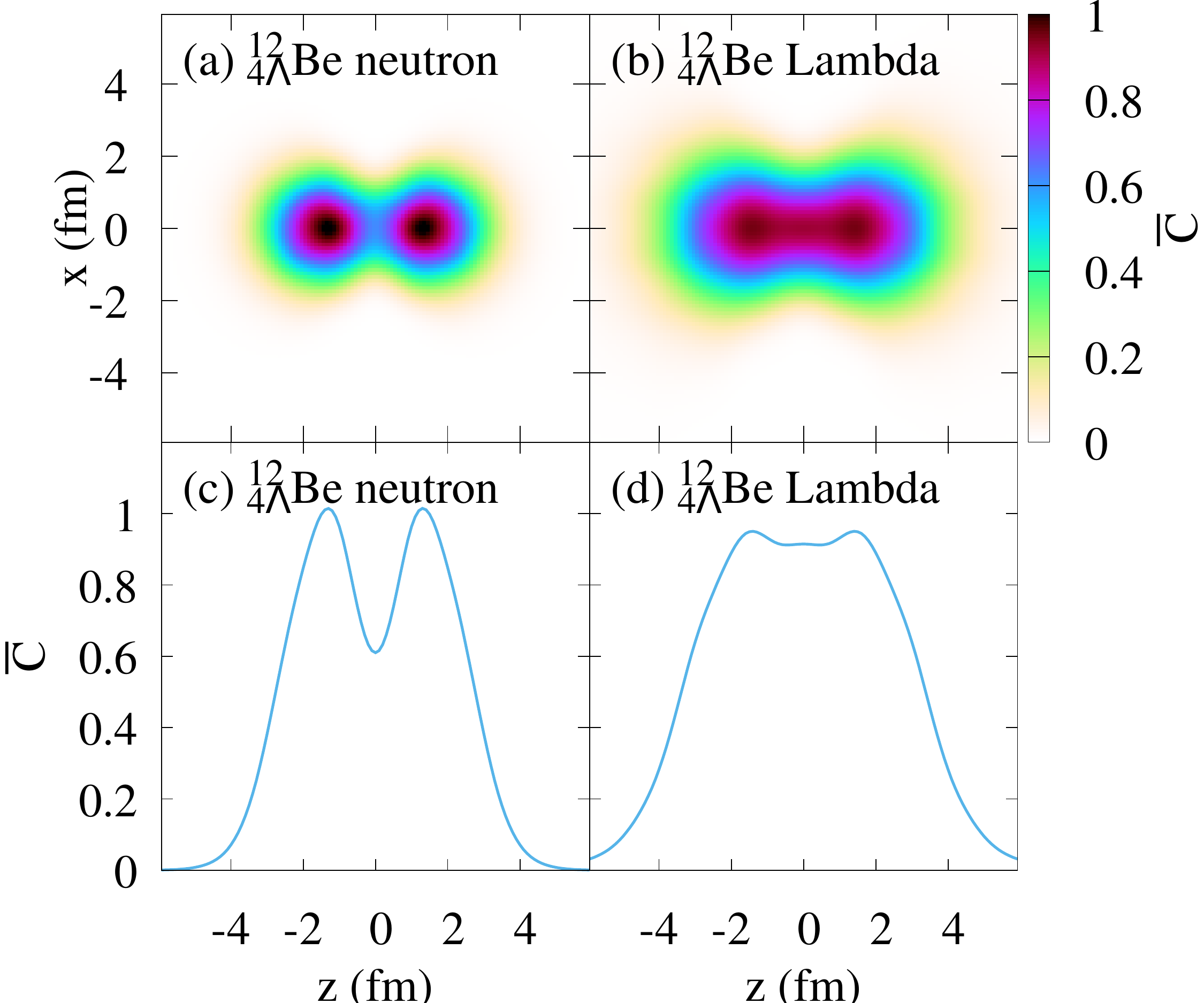}
\end{center}
\caption{Localization functions 
$\overline{\cal C}_{q\sigma} = {\cal C}_{q\sigma}
\frac{\rho_{q\sigma}}{{\rm max}\rho_{q\sigma}}$ 
on $zx$ plane in $^{12}_{4\Lambda}$Be
for (a) $(q\sigma)=({\rm neutron}\uparrow)$ and (b) $(q\sigma)=(\Lambda\uparrow)$. 
In panels (c) and (d) are plotted $\overline{\cal C}_{n\uparrow}$ and 
$\overline{\cal C}_{\Lambda\uparrow}$, respectively, along the symmetry axis. }
\label{fig:nlf2d_be}
\end{figure}

In Fig. \ref{fig:nlf2d_be}, the neutron and $\Lambda$ hyperon localization measures, 
$\overline{\cal C}_{n\uparrow}$ and $\overline{\cal C}_{\Lambda\uparrow}$ defined by 
Eq. (\ref{eq:Cbar}) are plotted. 
The neutron localization has two strong peaks at the positions of the two $\alpha$ clusters, 
indicating the $\alpha$-cluster correlation. 
More interestingly, the localization of $\Lambda$ hyperon in Fig. \ref{fig:nlf2d_be} (b) and (d) 
takes the values slightly more around the $\alpha$ clusters than at the region in between, 
which was not seen in the one-body density distribution in Fig. \ref{fig:den1d_be}. 
This implies the existence of two $^6_{2\Lambda}$He clusters, 
although the localization of the $\Lambda$ hyperons is much more obscure than the nucleon. 

We remark also that the rms radius of $\Lambda$ hyperon in $^{12}_{4\Lambda}$Be 
is significantly larger than that of nucleon ($R_\Lambda/R_N=3.49/2.42\approx 1.4$) 
although the number of $\Lambda$ particles in $^{12}_{4\Lambda}$Be
is the same as those of neutrons and protons. 
This is because the $\Lambda$ particles are more weekly bound in a shallow potential ($\approx -30$ MeV) 
as one sees in Fig. \ref{fig:pot_be} (b). Especially, the last occupied orbital in$^{12}_{4\Lambda}$Be 
is bound only by 1.3 MeV, which makes a halo- or skin-like structure of $\Lambda$. 
Similar results have been obtained also in Refs. \cite{Rufa90,Sch92,LM02} for spherical multi-$\Lambda$ systems.

\section{Results and discussion: Ne and Si hyper isotopes}\label{sec:results2}

Next, we investigate Ne and Si hyper isotopes with two and eight $\Lambda$ particles: 
$^{20}$Ne, $^{22}_{2\Lambda}$Ne, $^{28}_{8\Lambda}$Ne, 
$^{28}$Si, $^{30}_{2\Lambda}$Si, and $^{36}_{8\Lambda}$Si. 

\begin{figure}
\begin{center}
\includegraphics[width=8.5cm]{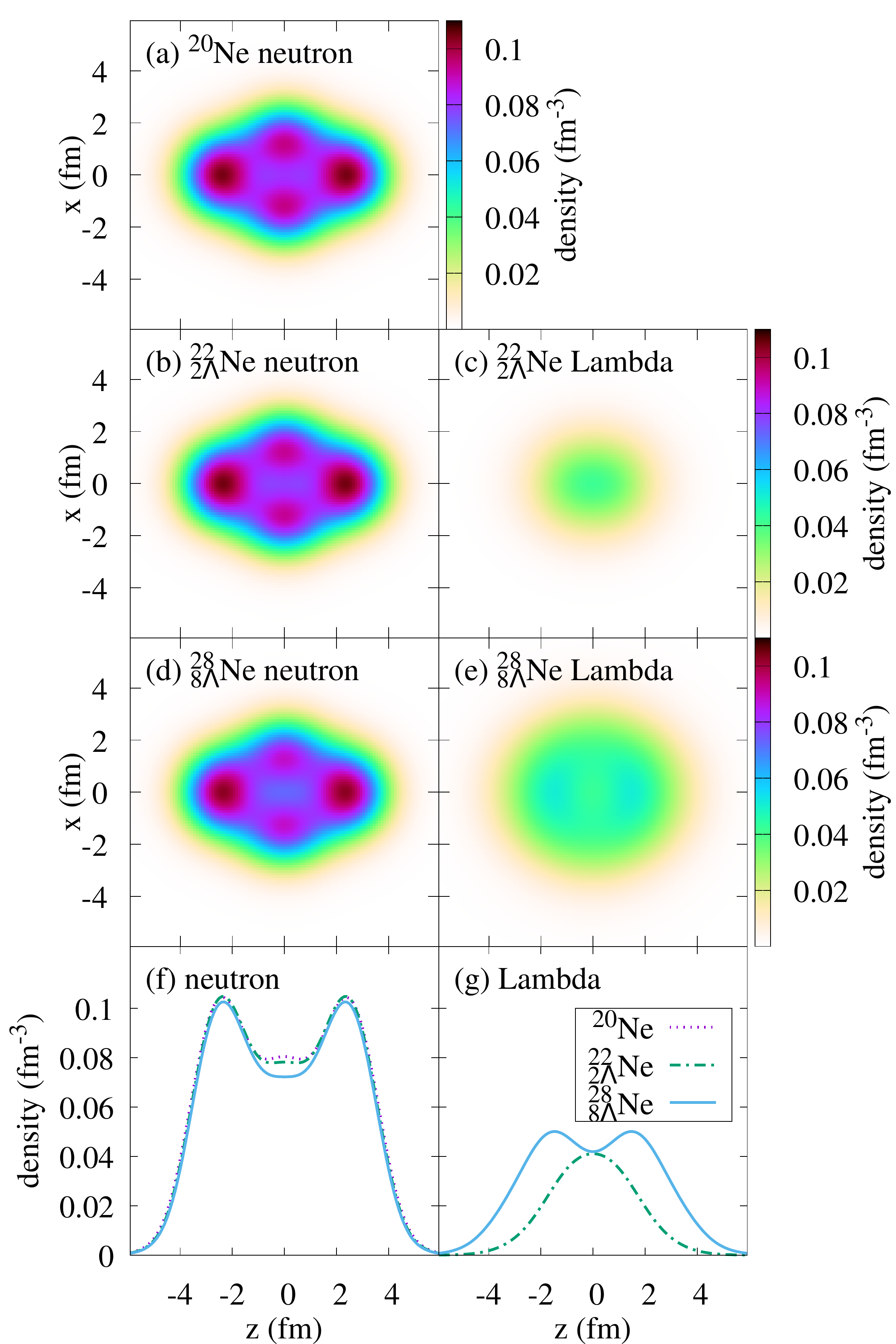}
\end{center}
\caption{The density distributions on $zx$ plane in the Ne hyper isotopes, 
(a) neutron in $^{20}$Ne, 
(b) neutron in $^{22}_{2\Lambda}$Ne, (c) $\Lambda$ hyperon in $^{22}_{2\Lambda}$Ne, 
(d) neutron in $^{28}_{8\Lambda}$Ne, (e) $\Lambda$ hyperon in $^{28}_{8\Lambda}$Ne. 
Panels (f) and (g) shows the comparisons of neutron and $\Lambda$ densities along the symmetry axis, 
respectively among the hyper isotopes. 
The $z$ axis (horizontal) is the symmetry axis. }
\label{fig:den2d_ne}
\end{figure}

In Fig. \ref{fig:den2d_ne} we show the density distributions of neutron and $\Lambda$ hyperon in $^{20}$Ne, $^{22}_{2\Lambda}$Ne, and $^{28}_{8\Lambda}$Ne. 
In panels (a)-(e) of Fig. \ref{fig:den2d_ne}, the density distributions 
of individual hyper isotopes on $zx$ plane are plotted, and in panels (f) and (g), 
the densities of different hyper isotopes along the symmetry axis are compared. 
Notice that the axis of symmetry is horizontal in Fig. \ref{fig:den2d_ne}. 
The quadrupole deformation parameters $\beta_2$ and radii $R$ are tabulated in Table \ref{tb:q2r_nesi}. 

The nucleon in the Ne hyper isotopes has prolate deformation. 
The density distributions of $\Lambda$ particles are also prolately deformed so that 
they gain binding energy from the attractive interaction with nucleons. 
From Table \ref{tb:q2r_nesi}, we see that 
the nucleon quadrupole deformation $\beta_{2N}$ decreases from $^{20}$Ne to $^{22}_{2\Lambda}$Ne 
and from $^{22}_{2\Lambda}$Ne to $^{28}_{8\Lambda}$Ne. 
The deformation of $\Lambda$ hyperon $\beta_{2\Lambda}$ also decreases from the $2\Lambda$ to the $8\Lambda$ isotopes. 

\begin{figure}
\begin{center}
\includegraphics[width=8.5cm]{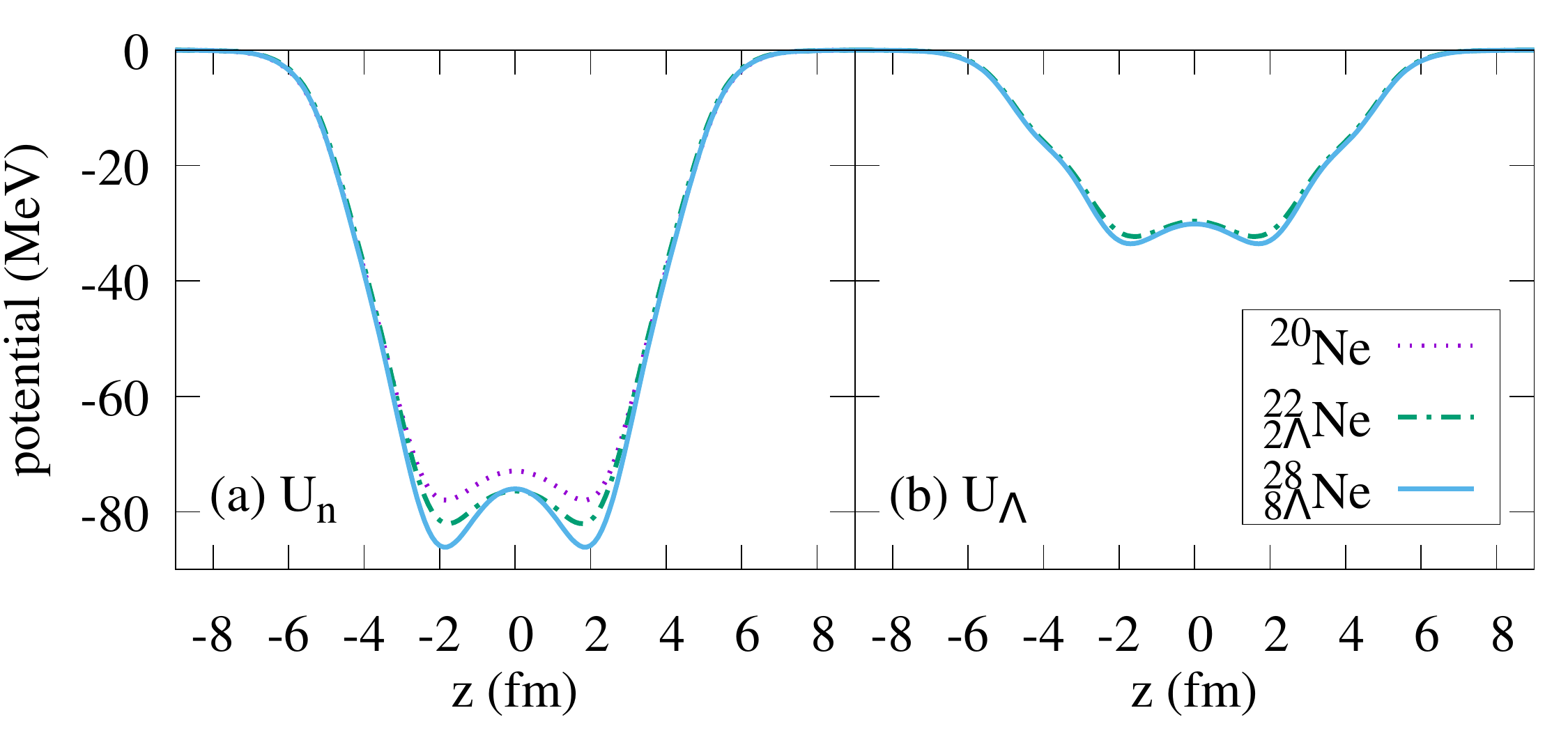}
\end{center}
\caption{Central potentials of (a) neutron and (b) $\Lambda$ hyperon of Ne hyper isotopes, 
$^{20}$Ne, $^{22}_{2\Lambda}$Ne, $^{28}_{8\Lambda}$Ne. }
\label{fig:pot_ne}
\end{figure}

We also observe in the density distributions that the hole at the center of neutron density 
becomes significantly depressed when there are eight $\Lambda$ hyperons. 
In Fig. \ref{fig:pot_ne} are shown the central mean fields of neutron and $\Lambda$ hyperon 
in the Ne hyper isotopes. 
As seen from the difference between the dotted and the dot-dashed curves 
in Fig. \ref{fig:pot_ne} (a), in $^{22}_{2\Lambda}$Ne, 
the $\Lambda$ particles in the lowest $s$-shell orbital give an extra contribution 
to the central potential that shifts down the bottom of it. 
When six more $\Lambda$ particles are added ($^{28}_{8\Lambda}$Ne), they occupy the 
$p$-shell orbitals whose density is vanishing at the origin [compare the solid and dot-dashed curves 
in Fig. \ref{fig:den2d_ne} (g)]. 
Consequently, the mean-field potential shown with the solid curve in Fig. \ref{fig:pot_ne} (a) 
is deepened at regions where $z\approx\pm 2$ fm. The similar thing happens in the other directions as well. 
This change in the potential induces the deeper hole in the nucleon density of $^{28}_{8\Lambda}$Ne. 

\begin{table}
\caption{Quadrupole deformation parameters $\beta_{2}$ and root-mean-squared radii $R$ of 
the Ne and Si hyper isotope chains. $\beta_{2N}$ and $R_N$ are deformation parameter and radius 
calculated from the nucleon density, 
and $\beta_{2\Lambda}$ and $R_\Lambda$ are calculated from the $\Lambda$ hyperon density. 
$N_\Lambda$ is the number of $\Lambda$ hyperons. }
\begin{tabular}{ccccc}
\hline\hline
$N_\Lambda$ & $\beta_{2N}$ & $\beta_{2\Lambda}$ & $R_N$ (fm) & $R_\Lambda$ (fm) \\
\hline
\multicolumn{5}{c}{Ne hyper isotopes} \\
$0$ &  $0.43$ & $-$       & $2.82$ & $-$ \\
$2$ &  $0.40$ & $0.16$ & $2.81$ & $2.53$ \\
$8$ &  $0.38$ & $0.09$ & $2.82$ & $3.33$ \\
\multicolumn{5}{c}{Si hyper isotopes} \\
$0$ &  $-0.31$ & $-$       & $3.14$ & $-$ \\
$2$ &  $-0.22$ & $-0.12$ & $2.95$ & $2.49$ \\
$8$ &  $0.00$ & $0.00$ & $2.92$ & $3.14$ \\
\hline\hline
\end{tabular}
\label{tb:q2r_nesi}
\end{table}

\begin{figure}[h]
\begin{center}
\includegraphics[width=8.5cm]{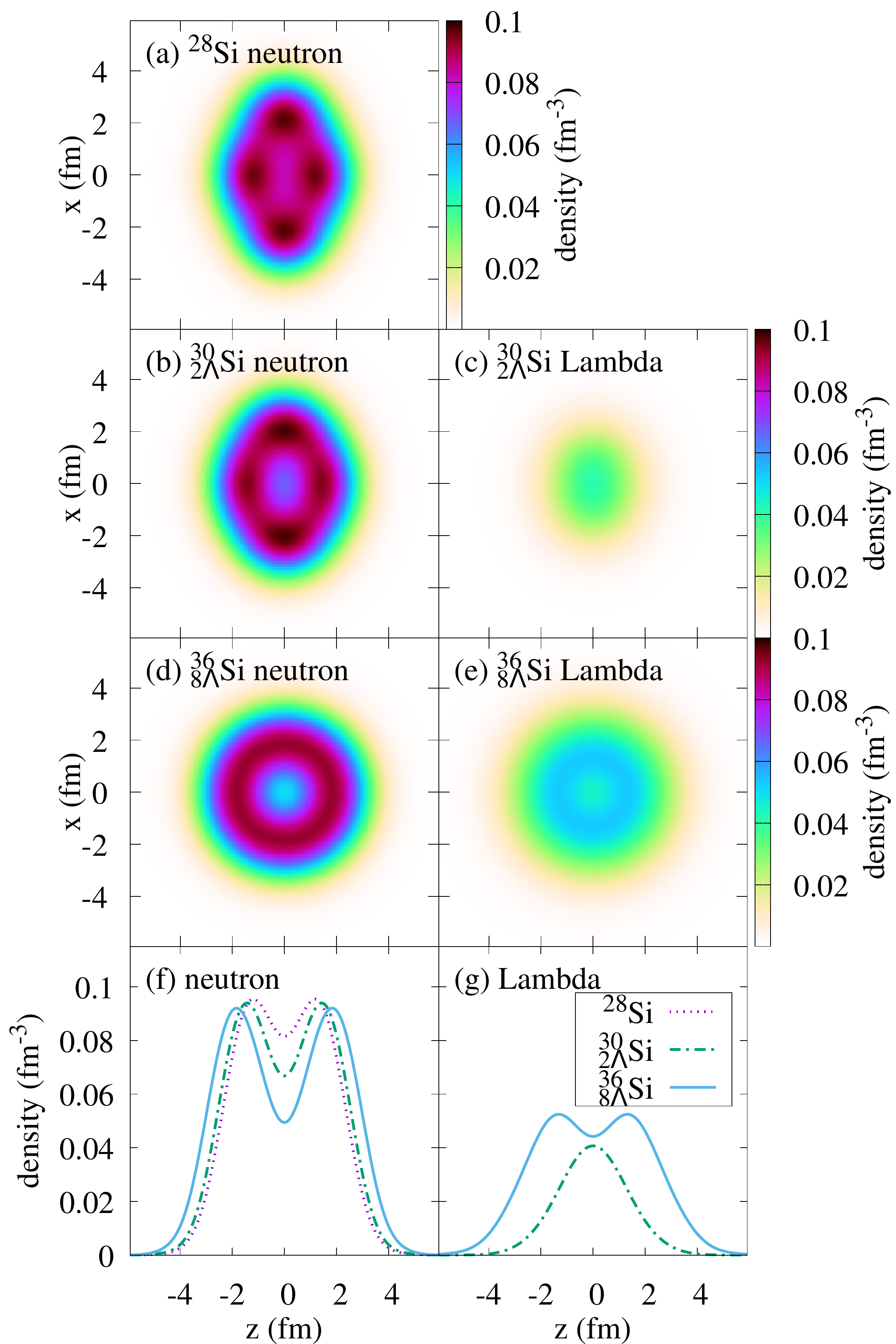}
\end{center}
\caption{Neutron and $\Lambda$ hyperon density distributions of Si hyper isotopes, 
$^{28}$Si, $^{30}_{2\Lambda}$Si, and $^{36}_{8\Lambda}$Si. 
The $z$ axis (horizontal) is the symmetry axis. }
\label{fig:den2d_si}
\end{figure}

In the Si hyper isotope chain, the shape variation is more drastic. 
In Fig. \ref{fig:den2d_si}, we show the density distributions of neutron and $\Lambda$ hyperon in $^{28}$Si, 
$^{30}_{2\Lambda}$Si, and $^{36}_{8\Lambda}$Si, and the quadrupole deformations and radii are 
tabulated in Table \ref{tb:q2r_nesi}. 
$^{28}$Si is obtaletly deformed. 
With two $\Lambda$ hyperons, the oblate deformation is relaxed from $\beta_{2N}=-0.31$ to $-0.22$. 
With eight $\Lambda$ hyperons, the spherical magicity of $\Lambda$ overcomes 
the nucleon deformation, and the system becomes spherical.  
The nucleons are in $(1s_{1/2})^4(1p_{3/2})^8(1p_{1/2})^4(1d_{5/2})^{12}$ configuration 
in $^{32}_{8\Lambda}$Si. 
Similar results are obtained also in the previous works on single-$\Lambda$ hyper nuclei \cite{MH08,Lu11}. 
Since the potential energy surface of $^{28}$Si is rather flat as a function of the quadrupole deformation, 
the Si hyper isotopes with one, two, or eight $\Lambda$ hyperons, which resist against deformation, 
easily go towards a spherical shape \cite{MH08,Lu11}. 

As in the case of the Be hyper isotopes, the radius of $\Lambda$ hyperon is larger than that of nucleon 
when $\Lambda$'s are in the $p$-shell which are more loosely bound than the nucleon 
[see Table \ref{tb:q2r_nesi} and Fig. \ref{fig:pot_ne} (b)].

\section{Summary and perspectives}\label{sec:summary}

The scope of hypernuclear physics is now being extended to multi-strangeness systems 
both in theoretical and experimental sides. 
In this work, we have investigated the clusterization and deformation properties of $N=Z$ multi-$\Lambda$ nuclei, 
$^{8+n}_{\ \ n\Lambda}{\rm Be}$, $^{20+n}_{\ \ \ n\Lambda}{\rm Ne}$, 
and $^{28+n}_{\ \ \ n\Lambda}{\rm Si}$, 
where $n = 2$, $4$ for Be, and  $n = 2$, $8$ for Ne and Si. 

We employed a relativistic mean-field model with meson-exchange interaction. 
One of the three coupling constants of the $\Lambda$-meson couplings is 
fitted to the binding energy of a single-$\Lambda$ hypernucleus $^{40}_\Lambda$Ca, 
and the remaining two are fixed by quark models. 

In $^{10}_{2\Lambda}$Be nucleus, when the two $\Lambda$'s occupy the $s$ ($p$) orbital, 
the intercluster distance decreases (increases), as in the case of $^9_\Lambda$Be \cite{Is11}. 

In $^{12}_{4\Lambda}$Be nucleus, a sign of two-$^{6}_{2\Lambda}$He cluster structure is observed 
in the two-body correlation embedded in the localization function, 
being consistent with the result obtained in Ref. \cite{MIB83}. 
Notice that, to draw a conclusion on the cluster correlation, one needs to perform a beyond-mean-field calculation which properly takes into account the quantum dynamics of shape degrees of freedom, 
and to analyze the spatial and spin-isospin correlations in the many-body wave function. 

When $\Lambda$ hyperons fill up the spherical major shells ($s$- and $p$-shells in the present study) 
but nucleons favor deformation, there is a competition between them 
due to the attractive interaction between the nucleons and the $\Lambda$ hyperons. 
In the Ne hyper isotopes $^{22}_{2\Lambda}$Ne and $^{28}_{8\Lambda}$Ne, 
the nucleon density distributions are slightly less deformed than in the normal isotope $^{20}$Ne, 
attracted by the $\Lambda$ hyperons that prefer a spherical shape. 
At the same time, the $\Lambda$ hyperon densities get deformed, attracted by the nucleons 
that prefer a deformed shape. 
The Si hyper isotopes are softer against the deformation and its relaxation. 
The deformation of the Si hyper isotopes changes more drastically as a function of the $\Lambda$ hyperon number. 

It is also seen that, 
as $\Lambda$ hyperons in the $p$-shell orbitals are more weekly bound than the nucleons, 
their density spreads out more than that of nucleon in the deformed systems as well as in spherical systems \cite{Rufa90, MZ93,Sch92,LM02}. 

There are several directions for future works to explore the generalized nuclear chart with 
the extra dimensions of hyperons. 
More systematic investigations on the ground-state structure of multi-$\Lambda$ nuclei would be interesting. 
For this purpose, it may be desirable to include $\Lambda$-$\Lambda$ and nucleon-$\Lambda$ 
pairing correlations as well as nucleon-nucleon pairing. 
Other interesting subjects are nuclei with many $\Sigma$ and/or $\Xi$ hyperons as well as $\Lambda$ hyperon. An extension of our model and numerical code to such systems is straightforward 
as long as the interaction is known. 
To further deepen our knowledge of the general multi-strangeness system, 
collective excitations of multi-strangeness nuclei should also be studied with, {\it e.g.}, 
GCM (geberator coordinate method) or RPA (random phase approximation) calculations.

\begin{acknowledgments}
The author thanks Kouichi Hagino for helpful discussions and comments on the manuscript. 
\end{acknowledgments}

\end{document}